\documentclass[12pt]{article}
\pdfoutput=1

\usepackage{putex}
\usepackage{amsmath,amssymb,amsfonts}
\usepackage{psfrag}
\usepackage{footmisc}
\usepackage{url}
\usepackage{mathtools}
\usepackage{color}

\usepackage{tikz}
    \usepackage{amssymb,amsfonts,amsmath}
    \usepackage{tkz-euclide}
        \usetikzlibrary{arrows,calc,patterns}
\usepackage{pgfplots}

\definecolor{darkblue}{rgb}{0.1,0.1,.7}
\definecolor{purple}{rgb}{0.6,0,0.6}
\definecolor{orange}{rgb}{0.9,0.6,0}
\usepackage[colorlinks, linkcolor=darkblue, citecolor=darkblue, urlcolor=darkblue, linktocpage]{hyperref} 
\usepackage[square, comma, compress,numbers]{natbib}
\usepackage[]{amsmath}
\usepackage[]{graphicx}
\usepackage[]{latexsym}
\usepackage[utf8]{inputenc}
\usepackage{geometry}
\usepackage{amscd}
\usepackage[all,cmtip]{xy}
\usepackage{mathrsfs}
\usepackage[margin=10pt,font=small,labelfont=bf]{caption}
\usepackage{dsdshorthand}
\usepackage{changepage}
\usepackage{setspace}
\usepackage{tikz}
\usepackage{subcaption}

\usepackage[titles]{tocloft}
\usetikzlibrary{arrows,positioning}

\def\SL2{\widetilde{SL}(2,\mathbb R)}

\numberwithin{equation}{section}

\newcommand {\beq} {\begin{equation*}}
\newcommand {\eeq} {\end{equation*}}
\newcommand {\bea} {\begin{eqnarray}}
\newcommand {\eea} {\end{eqnarray}}

\numberwithin{equation}{section}

\def\<{\langle}
\def\>{\rangle}

\tikzset{
    >=stealth',
    punkt/.style={
           rectangle,
           rounded corners,
           draw=black, very thick,
           text width=15em,
           minimum height=2em,
           text centered},
    pil/.style={
           ->,
           thick,
           shorten <=2pt,
           shorten >=2pt,}
}

 \def\ie{\begin{equation}\begin{aligned}}
\def\fe{\end{aligned}\end{equation}}
\begin{document}


\institution{IIT}{${}^1$ School of Basic Sciences, Indian Institute of Technology Bhubaneswar}
\institution{IIT}{
\qquad \qquad \qquad \qquad Bhubaneswar, Odisha, 752050, India}
\institution{CEICO}{${}^2$ Institute of Physics of the Czech Academy of Sciences \& CEICO}
\institution{CEICO}{ \qquad \qquad \quad Na Slovance 2, 18221 Prague, Czech Republic}

\title{ \bf
Thermodynamic curvature of charged black holes with $AdS_2$ horizons
}

\authors{Aditya Singh${}^{1}$,  Poulami Mukherjee${}^{1}$ and Chandrasekhar Bhamidipati${}^{1,2}$}

\abstract{
Sign and magnitude of the thermodynamic curvature provides empirical information about the nature of microstructures of a general thermodynamic system. For charged black holes in AdS, thermodynamic curvature is positive for large charge or chemical potential, and diverges for extremal black holes, indicating strongly repulsive nature. We compute the thermodynamic curvature at low temperatures, for charged black holes with AdS$_2$ near horizon geometry, and containing a zero temperature horizon radius $r_h$, in a spacetime which asymptotically approaches $AdS_D$ (for $D>3$). In the semi-classical analysis at low temperatures, the curvature shows a novel crossover from negative to positive side, indicating the shift from attraction to repulsion dominated regime near $T=0$, before diverging as $1/(\gamma T)$, where $\gamma$ is the coefficient of leading low temperature correction to entropy.  Accounting for quantum fluctuations, the curvature computed in the canonical ensemble is positive, whereas the one in the grand canonical ensemble, continues to show a crossover from negative to positive side. Moreover, the divergence of curvature at $T=0$ is cured irrespective of the ensemble used, resulting in a universal constant.
}
\date{}

\maketitle

\newpage
\setcounter{footnote}{0}
\noindent

\baselineskip 15pt

\section{Introduction}

Thermodynamics of charged black holes in AdS$_{D}$ spacetimes has been well studied, particularly with the motivation of understanding holographic field theories at finite temperature~\cite{Maldacena:1997re}-\cite{Myers99}. The presence of an AdS$_2$ factor in the near horizon geometry which has the form  AdS$_2 \times M_{d}$ (where $M_{d}$ is a compact space with $D=d+2$) is expected to give universal information about the low energy quantum theories for these black holes. The study of low energy aspects based on the AdS$_2$ factor has improved our understanding with the advent of Sachdev-Ye-Kitaev models \cite{SY92,GPS01,SS15,kitaev2015talk} and related studies in the low energy limit of string theories~\cite{Sen05}-\cite{H:2023qko}.\\

\noindent
A universal property of black holes in AdS$_{D}$ with charge $Q$ (and also SYK models) is that, there are interesting crossovers at low temperatures, with various thermodynamic quantities receiving non-trivial classical and quantum corrections~\cite{Sachdev:2019bjn,Iliesiu:2020qvm,Heydeman:2020hhw,Boruch:2022tno}. Energy and charge fluctuations of the low temperature quantum theory lead to a Schwarzian action~\cite{Davison17,Sachdev:2019bjn}:
\bea
I [f, \phi] &=& - S_0 (Q) +  \frac{K}{2} \int_0^{1/T} d \tau (\partial_\tau \phi - i (2 \pi \mathcal{E} T) \partial_\tau f)^2 \\ \nonumber &~&~~~~~~~~~ -  \frac{\gamma}{4 \pi^2} \int_0^{1/T} d \tau \, \{ \tan (\pi T f(\tau)), \tau\}, \label{Schwarzian}
\eea
with the following notation for the Schwarzian
\bea
 \left\{ g(\tau), \tau \right\} \,= \, \frac{g'''}{g'} - \frac{3}{2} \, \left( \frac{g''}{g'} \right)^2 \,.
 \eea
The action is specified in terms of three parameters, namely $\gamma$, the compressibility given as 
\bea
K = \frac{d Q}{d \mu} \Bigr|_{T=0} \,,
\eea
and the electric field at the horizon $\mathcal{E}$, which can be computed as
\bea
\frac{dS_0 (Q)}{dQ} = 2 \pi \mathcal{E} \, .
\eea
In particular, the dimensionless parameter $\mathcal{E}$ appeared first in the works of Sen~\cite{Sen05} while proposing an entropy function for general black holes (see~\cite{GPS01,SS15} for its appearance in complex SYK models). Unlike the zero temperature entropy $S_0$ and $\mathcal{E}$, the compressibility $K$ and the coefficient $\gamma$ are not universal  and depend upon the UV details of the theory. There are further novel features to the above model with the inclusion of quantum corrections and also including supersymmetric situations~\cite{Iliesiu:2020qvm,Heydeman:2020hhw,Boruch:2022tno}\\

\noindent
The aim of this paper is to attempt an understanding of the nature of microstructures of such low temperature charged black holes in AdS following recent developments~\cite{Stanford:2017thb,Sachdev:2019bjn,Iliesiu:2020qvm,Heydeman:2020hhw,Boruch:2022tno,Geng:2022tfc}, using the methods of thermodynamic geometry. The main idea of this approach is based on thermodynamic fluctuation theory, which starts from writing the number of microstates of a thermodynamic system as
\begin{equation}\label{omega}
\Omega = e^{\frac{S}{k_B}}\, ,
\end{equation}
where $k_B$ is the Boltzmann constant. One now considers a thermodynamic system $I_0$ in equilibrium, with a  sub-system $I$ in it, in addition to having a couple of independent fluctuating variables, $x^i$ where $i=1,2$. The number of microstates in eqn. (\ref{omega}) can now be related to the probability $P(x^1,x^2)$ of locating the state of the system somewhere between $(x^1,x^2)$ and $(x^1 + dx^1,x^2+dx^2)$ . According to second law of thermodynamics the pair $(x^1,x^2)$  picks the values that maximise the entropy $S=S_{\text{max}}$. In effect, $(x^1,x^2)$ describe fluctuations around the maximum and the probability around this maximum can be written as ~\cite{Ruppeiner:1995zz}:
\begin{equation}
P(x^1,x^2) \propto e^{-\frac{1}{2}\, \Delta l^2} \, ,
\end{equation}
where the line element which measures thermodynamic distance between two nearby fluctuation states is written as:
\begin{equation}\label{distance1}
\Delta l^2 \, = -\frac{1}{k_B}\, \frac{\partial^2 S}{\partial x^i\partial x^j}\, \Delta x^i \Delta x^j\, .
\end{equation}
The distance is then shorter when the fluctuation between neighbouring states is more probable. The thermodynamic curvature $R$ computed from the line element in eqn. (\ref{distance}) contains much information, and has been studied for various systems in nature, such as Ideal/van der Waals fluids, quantum gases to other Bose/Fermi systems, including the Ising model and black holes~\cite{weinhold}-\cite{Bhattacharya:2017hfj}. The current understanding based on the available empirical data is that repulsive (attractive) interactions of a thermodynamic system turn out to have positive (negative) value of $R$~\cite{Ruppeiner:1995zz}.  $R$ typically diverges at the phase transitions points and has zero crossings at the points where the attractive and repulsive interactions are in balance or in a non-interacting situation. \\

\noindent
 Thermodynamic geometry of charged black holes in AdS has been well studied in the canonical, grand canonical and also other mixed ensembles in several works (see e.g.,~\cite{Sahay:2010tx,Niu:2011tb}). It has also been observed that the curvature $R$ is generally positive for the large black hole branch above a certain critical charge or chemical potential and diverges as $+1/T$ near $T=0$, i.e., in the extremal limit. These computations are done for charged black holes in AdS, far away from the horizon. In this work, out motivation is to study the computation of $R$ for charged black holes in AdS, in a certain low temperature regime where the dynamics is governed by corrections to the AdS$_2$ near horizon geometry. The partition function has been computed 
recently in both canonical and grand canonical ensembles, carefully taking into account the quantum fluctuations~\cite{Stanford:2017thb,Sachdev:2019bjn,Iliesiu:2020qvm,Heydeman:2020hhw,Boruch:2022tno,Geng:2022tfc}, resolving the mass-gap puzzle~\cite{Iliesiu:2020qvm,Heydeman:2020hhw}. As noted above, the thermodynamic variables of black holes in the low temperature limit show certain universal properties, but there are also non-universal behaviours, which depend on the how the AdS$_2$ is embedded in the higher dimensional theory. Here, universality of thermodynamic quantities implies independence of their characteristics on parameters of the geometry far from the AdS$_2$ near-horizon region. Indeed, the behaviour of $R$ in the low temperature limit we find here, is due to the presence of a near horizon AdS$_2$ geometry, and  its nature differs far away from the horizon. More importantly, $R$ diverges on the positive side at $T=0$, in the semi-classical limit, which is similar to earlier calculations done in the full geometry~\cite{Sahay:2010tx,Niu:2011tb,Ruppeiner:2023wkq}. In addition, the curvature contains a novel crossover from the  negative to positive side, which objectively points towards the existence of attractive interactions (bosonic in nature as per our conventions) which develop at low temperatures, i.e., for $T \ll1/r_h$, before becoming repulsive. Once the quantum corrections to thermodynamic quantities are included, both the curvatures computed in the canonical and grand canonical ensembles and evaluated at $T=0$, turn out to be equal to a constant. The key take away message is that the quantum corrections to the near horizon nearly AdS$_2$ geometry cure divergence of the thermodynamic curvature at $T=0$. This indicates that the microstructures are weakly repulsive in nature at $T=0$, as opposed to the strongly repulsive nature anticipated from the analysis done earlier in the full geometry~\cite{Sahay:2010tx,Niu:2011tb,Ruppeiner:2023wkq}. \\

\noindent
The rest of the paper is as structured as follows. In section-(\ref{2}), we set up the notations and collect known details of thermodynamic geometry of charged blackholes in AdS spacetimes. These computations will be for charged black holes in D-dimensions far from the horizon. In particular, we note from figure-(\ref{RhighT}), that the thermodynamic curvature is generally positive for large black holes above a certain chemical potential and diverges at $T=0$. In section-(\ref{3}), we perform a computation of the thermodynamic curvature in semi-classical as well as with the inclusion of quantum corrections to thermodynamic quantities. Subsection-(\ref{3.1}) contains the computation of thermodynamic curvature in the semi-classical limit, where the corrections to entropy come from corrections to the AdS$_2$ geometry~\cite{Sachdev:2019bjn} and show that there is a novel crossover. The $T=0$ behaviour is unmodified and matches with earlier results~\cite{Sahay:2010tx,Niu:2011tb,Ruppeiner:2023wkq}. In subsection-(\ref{3.2}), we follow~\cite{Iliesiu:2020qvm} and note down the thermodynamic quantities in the low temperature limit, where the system can also be described by an effective one dimensional Schwarzian action, based on corrections to the near horizon AdS$_2$ geometry. The quantum corrections to the thermodynamic quantities in the canonical and grand canonical ensembles are now known~\cite{Iliesiu:2020qvm}, and we use these to compute the curvature in both these cases. In the canonical ensemble, the curvature is generally positive, with no crossover and is a positive constant at $T=0$. In the grand canonical ensemble, the curvature is a positive constant at $T=0$, but crosses over to the negative side with slight increase of temperature, but still well below $1/r_h$. At $T=0$, both the curvatures are finite as well as independent of charge, giving  a universal constant. We end with remarks in section-(\ref{4}).

\section{Charged black holes in AdS$_D$ } \label{2}

\begin{figure}[h!]
 	{\centering
 		{\includegraphics[width=4in]{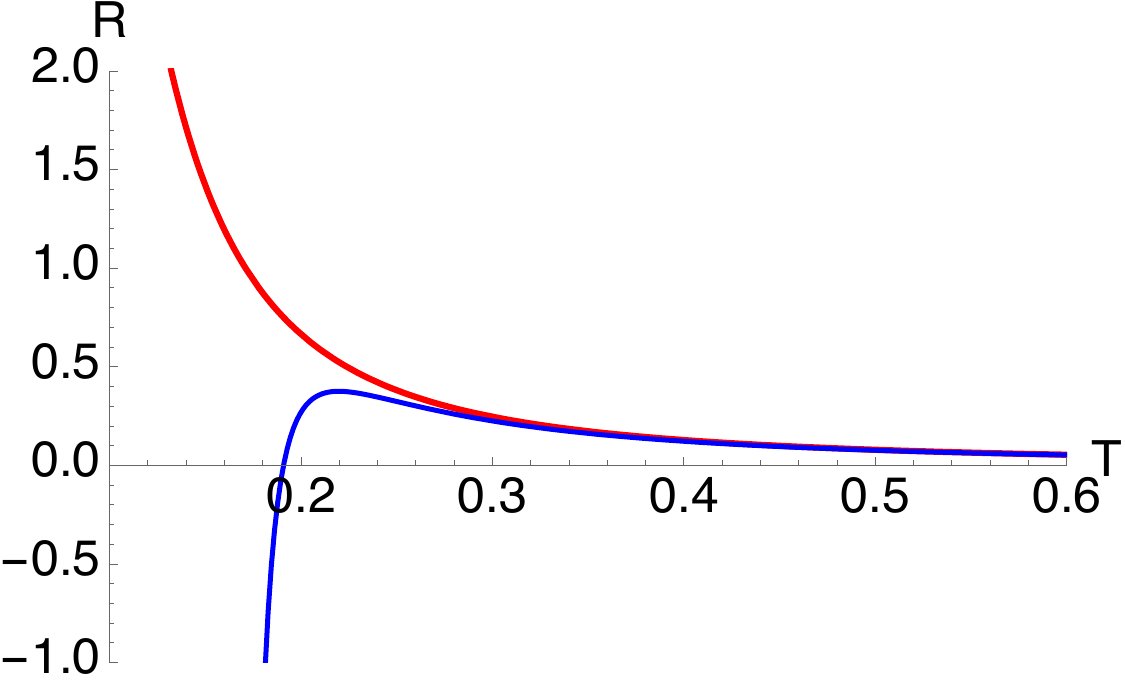}}\hspace{0.4cm}
 		
 		\caption{\footnotesize Thermodynamic curvature of charged black holes in AdS$_4$ in the $(T, \mu)$ plane) for $\mu >1$ (red curve) and $\mu <1$ (blue curve). The red curve is for $\mu = 1.05$ and the blue curve for $\mu = 0.8$.  \label{RhighT}	}  
 	}  
 \end{figure}

Let us start with the action of Einstein-Maxwell theory in AdS$_{d+2}$ ($d > 1$) in the presence of a U(1) gauge field with the action~ \cite{Myers99}.
\bea
I =  \int d^{d+2} x \sqrt{g} \left[ -\frac{1}{2 \kappa^2} \left(\mathcal{R}_{d+2} + \frac{d(d+1)}{L^2} \right) +  F^2 \right], \label{EM}
\eea
where we set $\kappa^2 = 8 \pi$ and the gravitational constant $G_N=1$.  $\mathcal{R}_{d+2}$ is the Ricci scalar, with 
$L$ denoting the AdS$_{d+2}$ radius. The above theory is known to contain black hole solutions with the metric
\bea
ds^2 =  f(r) d\tau^2  + \frac{dr^2}{f(r)} + r^2 d \Omega_d^2 \, , \label{s1}
\eea
where $d \Omega_d^2$ stands for the metric of the $d$- dimensional sphere, and
\bea
f(r) = 1 + \frac{r^2}{L^2} + \frac{q^2}{r^{2d-2}} - \frac{m}{r^{d-1}}.
\eea
As $r \rightarrow \infty$, the metric in Eq.~(\ref{s1}) goes over to AdS$_{d+2}$, whereas near the horizon the geometry is AdS$_2 \times $S$_{d}$. 
The grand canonical potential is given as
\bea
\Omega (T, \mu) = \frac{\omega_d r_0 (T, \mu)^{d-1}}{16\pi} \left(1 - \frac{r_0 (T, \mu)^2}{L^2} \right) - \frac{\omega_d (d-1) \mu^2 r_0 (T, \mu)^{d-1}}{2d}\,. \label{Omega}
\eea
from which the entropy and charge can be obtained to be
\bea
S(T, \mu) =\frac{\pi ^{\frac{d+1}{2}} r_0^d}{2 \Gamma
   \left(\frac{d+1}{2}\right)}, \label{s5}
\eea
\bea
Q(T, \mu) =\frac{(d-1) \pi ^{\frac{d-1}{2}} \mu  r_0^{d-1}}{2 \Gamma
   \left(\frac{d+1}{2}\right)} \,, \label{s6}
\eea
where $\omega_d= \frac{2 \pi ^{\frac{d+1}{2}}}{\Gamma \left(\frac{d+1}{2}\right)}$. Focussing on the large black hole branch, the specific heat can be written as,
\bea
C_{\mu}=\frac{d^2 \pi ^{\frac{d+3}{2}} L^2 T \left(\frac{\sqrt{d L^2 \left(d^3
   \left(2 \mu ^2-1\right)-2 d^2 \mu ^2+d \left(-2 \mu ^2+4 \pi ^2 L^2
   T^2+1\right)+2 \mu ^2\right)}+2 \pi  d L^2 T}{d
   (d+1)}\right)^d}{\Gamma \left(\frac{d+1}{2}\right) \sqrt{d L^2
   \left(d^3 \left(2 \mu ^2-1\right)-2 d^2 \mu ^2+d \left(-2 \mu ^2+4
   \pi ^2 L^2 T^2+1\right)+2 \mu ^2\right)}} \, ,
\eea
Black holes in the fixed potential ensemble are known to show distinct behaviour for $\mu <1$ and $\mu >1$~\cite{Myers99} and our interest is in the later regime where extremal limit can be taken.
Using the grand potential, a thermodynamic line element can be constructed as
\begin{equation}\label{distance}
\Delta l^2 \, = -\frac{1}{T}\, \frac{\partial^2 \Omega (T, \mu)}{\partial x^i\partial x^j}\, \Delta x^i \Delta x^j,
\end{equation}
where the fluctuation coordinates $x^i$ are chosen as $(T, \mu)$. The above line element is conformally related to the one in eqn. (\ref{distance1}). The curvature following from it can be computed straightforwardly~\cite{Sahay:2010tx,Niu:2011tb}, though the resulting expression is quite long to express here. We instead show the result in figure-(\ref{RhighT}), which matches with the earlier computations~\cite{Sahay:2010tx,Niu:2011tb}. \\
 \vskip 0.1cm
 \noindent
The low temperature behaviour for $\mu>1$ is given below which will be useful later:
\bea \label{RT}
R = \frac{3 \sqrt{3} \left(2 \mu ^2-1\right)}{4 \pi ^2 \left(\mu
   ^2-1\right)^{3/2} L^3 T}-\frac{3 \left(6 \mu ^2-1\right)}{4
   \left(\pi  \left(\mu ^2-1\right)^2 L^2\right)}+\frac{\left(4
   \sqrt{3} \mu ^2+\sqrt{3}\right) T}{2 \left(\mu ^2-1\right)^{5/2}
   L}+O\left(T^2\right) \, .
\eea
We note that $R$ is positive and diverges as $T\rightarrow 0$, indicating dominant repulsive type interactions of microstructures.

\section{Low temperature nearly AdS$_2$} \label{3}

At sufficiently low temperatures, non-constant modes on $S_{d}$ are not excited and certain universal features can be obtained. 
The expressions in the previous section were for general values of $T$ and $\mu$ and in this section, following~\cite{Sachdev:2019bjn}, the low temperature expressions (at fixed $\mu=\mu_0$) can be obtained by writing $r_0$ as
\bea
r_0  = r_h(\mu_0)   + \frac{2 \pi L^2}{d+1} T + \mathcal{O}(T^2) 
\eea
where 
\bea
r_h \equiv L \left[\frac{(d-1)( \mu_0^2 8\pi (d-1) - d)}{d(d+1)} \right]^{1/2}\,, \label{rh}
\eea
is the zero temperature horizon radius and $\mu_0 = \mu |_{T=0}$. The expansion in eqn. (\ref{rh}), allows writing zero temperature entropy as
\bea
S_0  = \frac{2 \pi \omega_d}{8\pi} \, r_h^d \,, \label{S0}
\eea
where $\omega_d$ is the area of d-dimensional unit sphere. Note that the the entropy $S_0$ is given terms of the area of the horizon in the d-dimensional geometry. Compressibility can be written as
\bea
K = \frac{(d-1) \omega_d r_h^{d-3} \left[ d(d+1) r_h^2
+ (d-1)^2 L^2 \right]}{(d+1)}\,. \label{defK}
\eea
Zero temperature charge can be expressed in terms of $\mu_0$ as
\bea \label{Q0}
Q  = \frac{\left(d^2-1\right) \pi ^{\frac{d-1}{2}} \mu _0
   \left(\frac{L \sqrt{d \left(d^2-1\right) \left(d \left(2 \mu
   _0^2-1\right)-2 \mu _0^2\right)}}{d (d+1)}\right){}^{d-1}}{2
   (d+1) \Gamma \left(\frac{d+1}{2}\right)}\, .
\eea
As suggested in~\cite{Sachdev:2019bjn}, the low temperature analysis can be done in the canonical ensemble at constant $Q$. 
\subsection{Semi-classical analysis} \label{3.1}
The free energy in the fixed charge ensemble  $F =\Omega + \mu\, Q$ gets corrections to order $T^2$~\cite{Hawking:1995ap,Sachdev:2019bjn}:
\bea \label{free}
F= \frac{\pi ^{\frac{d-1}{2}} r_h^d \left(\frac{d^2 r_h}{(d-1)
   L^2}+d \left(\frac{1}{r_h}-\frac{2 \pi ^2 L^2 r_h
  T^2}{(d-1)^2 L^2+d (d+1) r_h^2}\right)-2 \pi  T\right)}{4
   \Gamma \left(\frac{d+1}{2}\right)}\, .
\eea
To this order, the entropy is given as
\bea
S(Q,T \rightarrow 0) = S_0 (Q)  + \gamma \, T + \cdots\, , \label{ST}
\eea
where $S_0 (Q)$  is the zero temperature entropy and is non-zero. The coefficient $\gamma$ (which also gives the gap scale $M^{-1}_{SL(2)}$) can be deduced from the corrections to the near horizon AdS$_2$ geometry\cite{AAJP15,Alm2016fws,JMDS16b,kitaev2015talk} as
\bea \label{gamma}
\gamma = \frac{d \pi ^{\frac{d+3}{2}} L^2 r_h^{d+1}}{\Gamma
   \left(\frac{d+1}{2}\right) \left((d-1)^2 L^2+d (d+1)
   r_h^2\right)}\,.
\eea
Due to the above scaling of entropy with temperature as in eqn. (\ref{ST}), it was believed that the statistical description breaks down at temperatures lower than $1/\gamma$. The scale $M_{SL(2)}$ is now understood as the energy  at which the (approximate) near horizon conformal symmetry of AdS$_2$ is broken. This becomes evident once the quantum corrections are taken into account~\cite{Iliesiu:2020qvm}. For now, we continue with the semi-classical analysis and return to consider the quantum corrected thermodynamic quantities in the next subsection. The low temperature behaviour of the chemical potential $\mu$ at leading order can be written as~\cite{Sachdev:2019bjn}
\bea
\mu (T) &=& \mu_0 - 2 \pi \mathcal{E} T + \ldots\, ,  \label{muT}
\eea
where
\bea
2\pi\mathcal{E} = \frac{\sqrt{2} \pi  L r_h \sqrt{d \left((d-1) L^2+(d+1) r_h^2\right)}}{(d-1)^2 L^2+d (d+1)
   r_h^2}\, .
\eea
The second term in eqn. (\ref{muT}) is purely the contribution at the boundary of AdS$_2$ and its contribution at the AdS$_{d+2}$ boundary can be computed in the presence of time diffeomorphisms and gives the second term in eqn. (\ref{Schwarzian}). We will see in the following section that eqn. (\ref{muT}) gets corrected further due to quantum fluctuations and we will use the temperature dependent terms there to compute the curvature. While using zero temperature relations, we will continue to express the thermodynamic relations in terms of either $r_h$, $Q$ or $\mu_0$ as per convenience as all these are constants, related to each other. We can compute the thermodynamic curvature in both canonical and grand canonical ensembles, but we prefer the later here, for comparison with the results in the previous section. 
The particular length ratio $L/r_h$ is kept fixed as the temperature is lowered in the limit $T \rightarrow 0$. We will set $L=1$ in the plots.\\

\noindent
The curvature can be computed straightforwardly by evaluating the metric from eqn. (\ref{distance}), and the expression for the four dimensional case is
\bea\label{Rlow}
R = \frac{\sqrt{3} \left(3 \left(1-2 \mu_0 ^2\right)^4-4 \pi ^2
   \left(4 \mu_0 ^6+8 \mu_0 ^4-11 \mu_0 ^2-1\right) L^2 T^2\right)}{4
   \pi ^2 \left(\mu_0 ^2-1\right)^{3/2} \left(2 \mu_0 ^2-1\right)^3
   L^3 T}
\eea
The above expression has been written in terms of $\mu_0$ for convenience, and can be converted to the fluctuating variable $Q$ using eqn. (\ref{Q0}). The first few terms of eqn. (\ref{Rlow}) close to $T=0$ are
\bea \label{Rlow1}
R =\frac{3 \sqrt{3} \left(2 \mu_0 ^2-1\right)}{4 \pi ^2 \left(\mu_0
   ^2-1\right)^{3/2} L^3 T}-\frac{\sqrt{3} \left(4 \mu_0 ^6+8 \mu_0
   ^4-11 \mu_0 ^2-1\right) T}{\left(\mu_0 ^2-1\right)^{3/2} \left(2
   \mu_0 ^2-1\right)^3 L}+O\left(T^2\right)
\eea
 \begin{figure}[h!]
 	{\centering
 		{\includegraphics[width=4in]{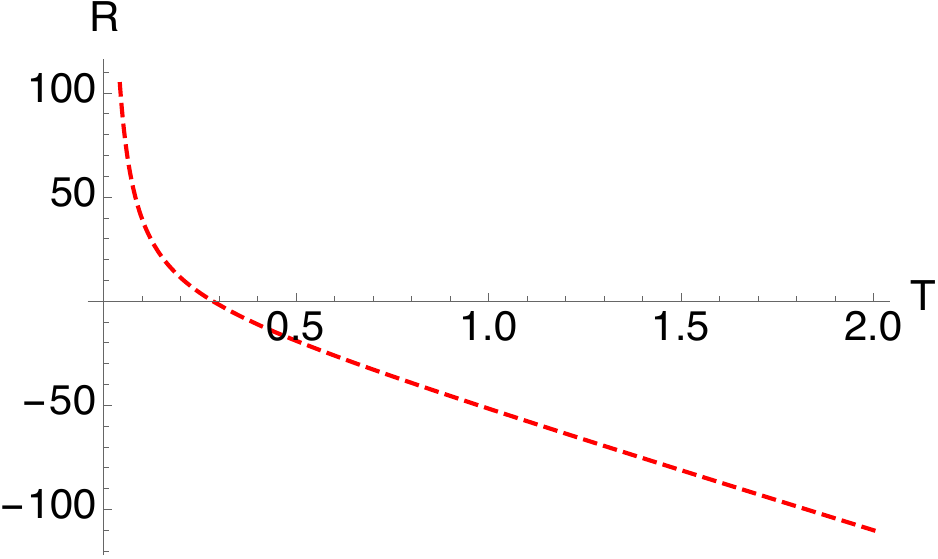}}\hspace{0.4cm}
 		
 		\caption{\footnotesize Thermodynamic curvature for charged black holes with near horizon AdS$_2$ geometry at low temperature. The curvature crosses over from negative to positive side at $T=0.282$ for $Q=0.2$ (or equivalently $\mu_0 = 1.052$). \label{RK}	}  
 	}  
 \end{figure}  
\vskip 0.1cm 
\noindent
The general result for curvature in figure-(\ref{RhighT}) is positive all along and approaches zero at high temperatures. However, the low temperature curvature in eqn. (\ref{Rlow}) behaves slightly differently, a seen in figure-(\ref{RK}). As the temperature is lowered (and still much below the energy scale $1/r_h$), the curvature can be negative, before resuming the positive behaviour and subsequent divergence as $T \rightarrow 0$. The $T=0$ behaviour is universal. That is, the first term of thermodynamic curvature in the general expression in eqn. (\ref{RhighT}) and the low temperature computation in eqn. (\ref{Rlow1}) agree exactly, showing the typical $1/T$ divergence in the extremal limit. In fact, we note that the coefficient of leading $1/T$ term in either of these low temperature series for curvature is exactly $1/\gamma$. \\
 \begin{figure}[h!]
 	{\centering
 		{\includegraphics[width=4in]{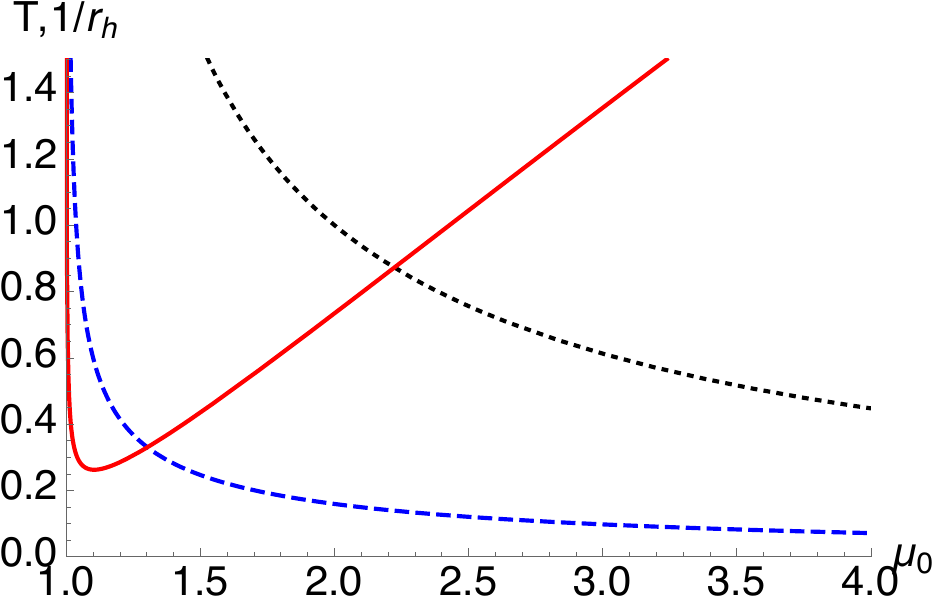}}\hspace{0.4cm}
 		
 		\caption{\footnotesize Red curve represents the zero crossings of curvature (eqn. (\ref{Zero})) as a function of $\mu_0$. The black dotted and blue dashed curves represent respectively $1/r_h$ and $1/(2\pi r_h)$ plotted w.r.t. to $\mu_0$. $d=2$ for all the cases. 
		 \label{Rzero}	}  
 	}  
 \end{figure}  
\vskip 0.1cm 
\noindent
Based on the empirical understanding of thermodynamic curvature, the points of crossover of $R$ indicate shift of the nature of collective interactions of microstructures from attraction dominated to repulsion dominated at a temperature, which is in the physical region of interest. In this same temperature regime, the black hole also shows a crossover and the description in terms of an effective action governed by the Schwarzian is feasible 
(see figure-2 in~\cite{Sachdev:2019bjn}). The zero crossing of the thermodynamic curvature in eqn. (\ref{Rlow}) occurs at
\bea \label{Zero}
T_{R=0} &=& \frac{\sqrt{3} \left(1-2 \mu_0 ^2\right)^2}{2 \pi  \sqrt{4 \mu_0 ^6+8 \mu_0 ^4-11 \mu_0 ^2-1} L}\, 
\eea
as also seen in figure-(\ref{Rzero}), happening at temperatures much lower than $1/r_h$. 
Of course, the crossing of $R$ shown in figure-(\ref{Rzero}) cannot be trusted at temperature comparable to $1/r_h$.

\subsection{Quantum corrections} \label{3.2}

In this subsection, we follow the analysis in~\cite{Iliesiu:2020qvm}, where the corrections to thermodynamic quantities coming from quantum fluctuations were computed for Reissner-Nordstr\"{o}m near-extremal black holes at low temperatures. One of the key findings of~\cite{Iliesiu:2020qvm} is that in the fixed charge sector, the density of states of the system does not show a gap, but rather a continuum of states, for the case of non-supersymmetric black holes. The computed partition function in the canonical and grand canonical ensembles shows that the low energy limit is well captured by two dimensional Jackiw-Teitelboim gravity, which is coupled to a $U(1)$ gauge field and other gauge fields coming from dimensional reduction. We first compute the thermodynamic curvature in the canonical ensemble and then in the grand canonical ensemble. \\

\begin{figure}[h!]
 	{\centering
 		{\includegraphics[width=4in]{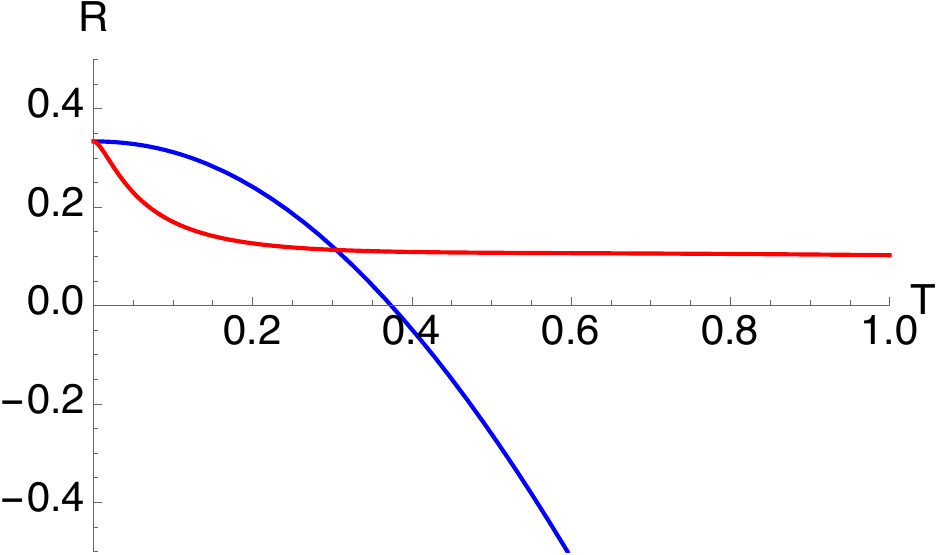}}\hspace{0.4cm}
 		
 		\caption{\footnotesize Red and blue curves represent respectively, the thermodynamic curvatures in canonical and grand canonical ensembles in four dimensions for $Q=2, L=1$. The value at $T=0$ is $R=1/3$.
		 \label{RC}	}  
 	}  
 \end{figure}

\begin{figure}[h!]
 	{\centering
 		{\includegraphics[width=4in]{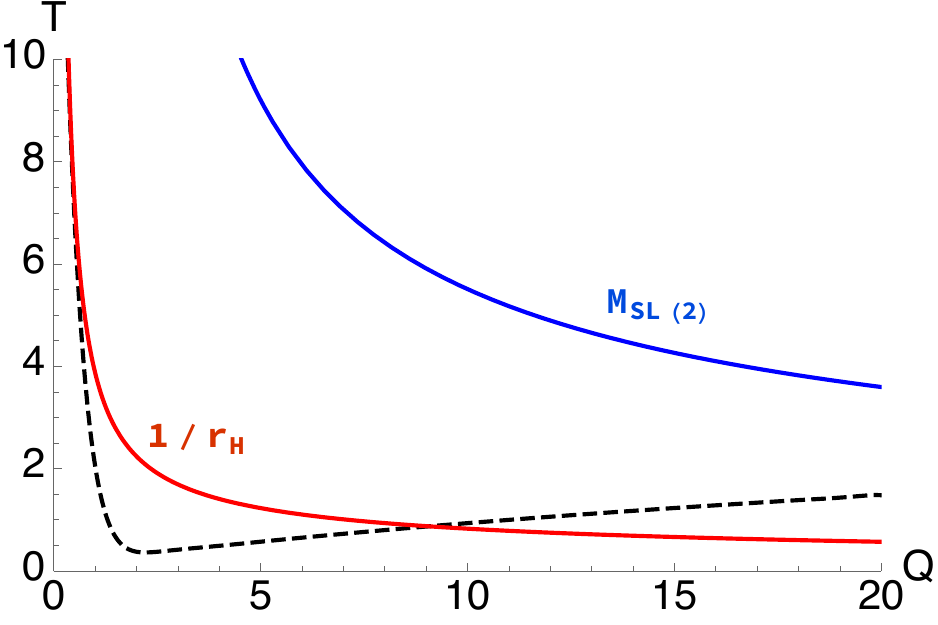}}\hspace{0.4cm}
 		
 		\caption{\footnotesize The dashed curve represents the temperature at which the curvature in the grand canonical ensemble has zeroes as a function of charge $Q$ (or equivalently the chemical potential $\mu_0$). The red and blue curves represent the inverse of horizon radius and the conformal symmetry breaking scale $M_{SL(2)}$ as a function of charge.
		 \label{RzeroC}	}  
 	}  
 \end{figure}

\noindent
The partition function in the fixed charge ensemble was shown in~\cite{Iliesiu:2020qvm} to be
\be 
\label{Qensemble}
Z_{\rm RN}[T,Q] = \Big( \gamma\, T\Big)^{3/2} e^{\pi r_h^2-M_0(Q)/T+ 2\pi^2 \gamma T}.
\ee 
For our purposes, the quantum corrected entropy is
\bea\label{S1}
S &=& S_0+ \gamma  T+\frac{3}{2} \log \left(\gamma  T\right)\, ,
\eea
where the various quantities appearing in the above equation can be found to be~\cite{Iliesiu:2020qvm}
\be\label{extremal}
Q^2 =4\pi \left(r_h^2 + \frac{3r_h^4}{L^2}\right) ,~~~~M_0 = r_h\left( 1 + \frac{2r_h^2}{L^2}\right) \, .
\ee 
In the above equations, zero temperature entropy $S_0$ and $\gamma$ can be obtained from equations (\ref{S0}) and (\ref{gamma}) respectively, by setting $d=2$.
The specific heat following from eqn. (\ref{S1}) is:
\be
C=\frac{1}{18} T \left(\frac{27 }{T}+\frac{2 \sqrt{6} \pi ^{7/4} L^{5/2} \left(\sqrt{\pi  L^2+3 Q^2}-\sqrt{\pi }
   L\right)^{3/2}}{\sqrt{\pi  L^2+3 Q^2}}\right) \, ,
\ee
with the chemical potential found to be
\bea \label{mu1}
&&\mu = -\frac{\sqrt{L^2+3 r_h^2}}{8 \sqrt{\pi } L r_h \left(L^2+6 r_h^2\right)^3} \left(24 \pi ^2 L^4 r_h^3 T^2 \left(L^2+2 r_h^2\right)+L^2 T \left(L^2+6 r_h^2\right)  \right. \nonumber \\
&&~~~~~~~~~~~~~~~~~~~~~\left. \left(L^2 \left(8 \pi  r_h^2+27\right)+   6 r_h^2 \left(8 \pi  r_h^2+9\right)\right)-4 r_h \left(L^2+6 r_h^2\right)^3 \right) \, .
\eea
The partition function in the fixed potential ensemble reads~\cite{Iliesiu:2020qvm}
\be 
Z_{\rm RN}[\b, \mu] =e^{\mu \frac{Q_0}{T}+S_0(Q_0)- M_0(Q_0)/T} \Big( \gamma T\Big)^{3/2} e^{2\pi^2 \gamma T} ~Z_{U(1)},
\ee 
where $Z_{U(1)}$ is the contribution of the $U(1)$ mode, which we ignore for now it does not induce any major contributions to the curvature to be discussed below. With the above partition functions in both ensembles, one can now do the computation of thermodynamic curvatures in both the canonical and the grand canonical ensembles. For the canonical ensemble, one use the metric
\bea \label{FQ}
  dl^2 &=& -\frac{1}{T}\,\frac{\partial^2 F}{\partial T^2} (dT)^2 + \frac{1}{T}\,\frac{\partial^2 F}{\partial Q^2} (dQ)^2 \,. 
\eea
whereas for the grand canonical, one can use the same metric noted earlier in eqn. (\ref{distance}). For the grand canonical ensemble, one remembers that $Q$ is taken to be function of only $\mu_0$, which is the zero temperature contribution from eqn. (\ref{mu1}).
The thermodynamic curvature can be computed analytically as such in both the ensembles as before,  but the expressions are too long and hence we present the results in figure-(\ref{RC}).
There are major changes due to quantum corrections. First, the thermodynamic curvature at $T=0$ does not diverge anymore, irrespective of the ensemble used, and actually gives $R_{\rm T=0}= 1/3$, which is a universal constant independent of charge. This is in stark contrast to the behaviour of thermodynamic curvature noted earlier for full AdS$_D$ background~\cite{Mirza:2007ev,Sahay:2010tx,Niu:2011tb,Ruppeiner:2023wkq} and also the semi-classical analysis in the nearly AdS$_2$ near horizon geometry found in the last section. From the existing results on Ruppeiner geometry, one concludes that a small positive constant value of thermodynamic curvature at $T=0$ implies the presence of weakly interacting repulsive nature of degrees of freedom. In the grand canonical ensemble, as seen from figure-(\ref{RzeroC}), the thermodynamic curvature continues to have a crossover behaviour (already seen in the semi-classical analysis in the last section), which is below the scale $1/r_h$, and also below the conformal symmetry breaking scale $M^{-1}_{SL(2)}$.
Due to the arguments in~\cite{Iliesiu:2020qvm} for non-supersymmetric black holes, the thermodynamic analysis in this limit is justified and one expects the crossover to have a physical interpretation.

\section{Conclusions}  \label{4}

Charged black holes in AdS have long been used as holographic models for strongly interacting quantum systems at finite density\cite{Hartnoll:2016apf}. Close to the boundary, the geometry asymptotes to AdS$_{D}$, where the usual AdS/CFT relations hold giving an understanding of bulk properties in terms of field theory in $D-1$ spacetime dimensions. The low temperature correlations (for $D>3$) however were linked to the presence of the near horizon AdS$_2$ geometry alone~\cite{Faulkner09}. More recently, following examples of the SYK model, interesting one dimensional Schwarzian action has been studied~\cite{JMDS16b,AAJP15,Alm2016fws,Davison17,Stanford:2017thb,GKST19,Liu:2019niv}, which has allowed novel computation of quantum properties, in comparison to the AdS$_D$ approach. The effective one dimensional action has also been obtained from the low energy limit of the Einstein-Maxwell theory which contains the charged black holes in asymptotically AdS$_{D}$ spacetime~\cite{Sachdev:2019bjn,Iliesiu:2020qvm,Heydeman:2020hhw,Boruch:2022tno}.  In particular, at low temperatures, the black holes exhibit interesting crossovers to the Schwarzian regime.The semi-classical analysis is now supplemented by the quantum corrections, which play a crucial role in resolving the mass gap puzzle in the non-supersymmetry near extremal black holes~\cite{Iliesiu:2020qvm}. \\

\noindent
An important byproduct of the recent developments in~\cite{Sachdev:2019bjn,Iliesiu:2020qvm,Heydeman:2020hhw,Boruch:2022tno} is that the thermodynamic quantities of charged black holes at low temperatures receive non-trivial corrections due to quantum fluctuations. Purely from thermodynamics point of view, Ruppeiner geometry is known to give reliable results on the nature of microscopic interactions of degrees of freedom, for a wide range of systems, starting from black holes to quantum gases. 
Thermodynamic geometry of charged black holes in AdS$_{D}$ studied earlier shows that the curvature is generally positive and diverges at $T=0$\cite{Sahay:2010tx,Niu:2011tb,Ruppeiner:2023wkq}. In this paper, we performed a computation of the curvature $R$ in the low temperature regime, both in the semi-classical as well as the quantum regime, based on the correction of thermodynamic quantities close to the near horizon geometry giving novel behaviour. In the semi-classical limit, as the temperature is lowered, keeping the charge fixed, $R$ shows crossover from negative (bosonic) to positive (fermionic) side (before diverging as $\frac{1}{\gamma T}$), in the region where the temperature is much less than $1/r_h$. Once the quantum corrections are taken into account, the curvature in the canonical ensemble is positive for all temperatures, but the one in the grand canonical ensemble continues to show a crossover for a wide range of charges and for $T \ll1/r_h$, which is also the region where the black hole undergoes interesting crossovers~\cite{Sachdev:2019bjn}. More importantly, both the curvatures are finite at $T=0$, equalling to a universal constant, independent of charge of black hole. This indicates that the interactions of microstructures are weakly repulsive, as opposed to the strongly repulsive nature anticipated from the analysis of thermodynamic curvature in the full AdS$_D$ geometry.\\

\noindent 
Some interesting questions which should be pursued are as follows. Black holes in higher derivative gravity are known to give non-trivial corrections to the entropy of black holes, both in supersymmetric and non-supersymmetric situations. It is important to study their thermodynamic geometry in the near horizon limit, with corrections to AdS$_2$ geometry as done here. Second, it has been suggested that the mass gap does exist for BPS black holes in supersymmetric theories~\cite{Heydeman:2020hhw,Boruch:2022tno} and it is interesting to check whether the thermodynamic curvature in either of the ensembles has any non-trivial behaviour beyond that found here.

\section*{Acknowledgements}
We thank Matthew Heydeman, Hao Geng and Sudipta Mukherji for helpful correspondence and comments on the draft.
A.S. wishes to thank the Council of Scientific and Industrial Research (CSIR), Government of India, for financial support. P.M. thanks IIT Bhubaneswar for Institute fellowship. C.B. thanks the DST (SERB), Government of India, for financial support through the Mathematical Research Impact Centric Support (MATRICS) grant no. MTR/2020/000135 and the Institute of Physics of the Czech Academy of Sciences \& CEICO, Prague, for warm hospitality. 
\bibliographystyle{apsrev4-1}
\bibliography{SRlow}
\end{document}